\documentclass[sigconf]{acmart}
\setcopyright{none}
\copyrightyear{2026}
\acmYear{2026}
\acmDOI{10.1145/3830397.3841867}

\usepackage{algorithm}
\usepackage{algpseudocode}
\usepackage{tabularx}
\usepackage{subcaption}

\AtBeginDocument{%
  }

\acmConference[UIST Adjunct '26]
{The 39th Annual ACM Symposium on User Interface Software and Technology Adjunct}{November 2--5, 2026}{Detroit, MI USA}
\acmISBN{979-8-4007-2855-6/26/11}

\begin{document}

\title{FocusAdapt: Context-aware Adaptive Focus Assistance in Diminished Reality }

\author{Tianyu Zhang}
\affiliation{%
  \institution{University of Rochester}
  \city{Rochester}
  \state{New York}
  \country{USA}
}
\email{tianyu.zhang@rochester.edu}

\author{Shutong Wu}
\affiliation{%
  \institution{University of Rochester}
  \city{Rochester}
  \state{New York}
  \country{USA}
}
\email{swu85@ur.rochester.edu}

\author{Jiankun Yang}
\affiliation{%
  \institution{University of Rochester}
  \city{Rochester}
  \state{New York}
  \country{USA}
}
\email{jyang118@u.rochester.edu}

\author{Zhen Bai}
\affiliation{%
  \institution{University of Rochester}
  \city{Rochester}
  \state{New York}
  \country{USA}
}
\email{zhen.bai@rochester.edu}

\author{Yukang Yan}
\affiliation{%
  \institution{University of Rochester}
  \city{Rochester}
  \state{New York}
  \country{USA}
}
\email{yukang.yan@rochester.edu}

\renewcommand{\shortauthors}{Trovato et al.}

\begin{abstract}
  Diminished Reality (DR) can reduce visual clutter by removing irrelevant objects. However, removing all task-irrelevant objects may eliminate useful contextual information and reduce situational awareness. We present FocusAdapt, a context-aware DR system that predicts object-level distraction by integrating visual saliency and similarity, semantic relevance, and gaze behavior. Based on findings from a formative study, FocusAdapt selectively diminishes highly distracting objects while preserving useful context, enabling adaptive focus assistance during procedural tasks.
\end{abstract}

\begin{teaserfigure}
  \begin{center}
    \includegraphics[width=\textwidth]{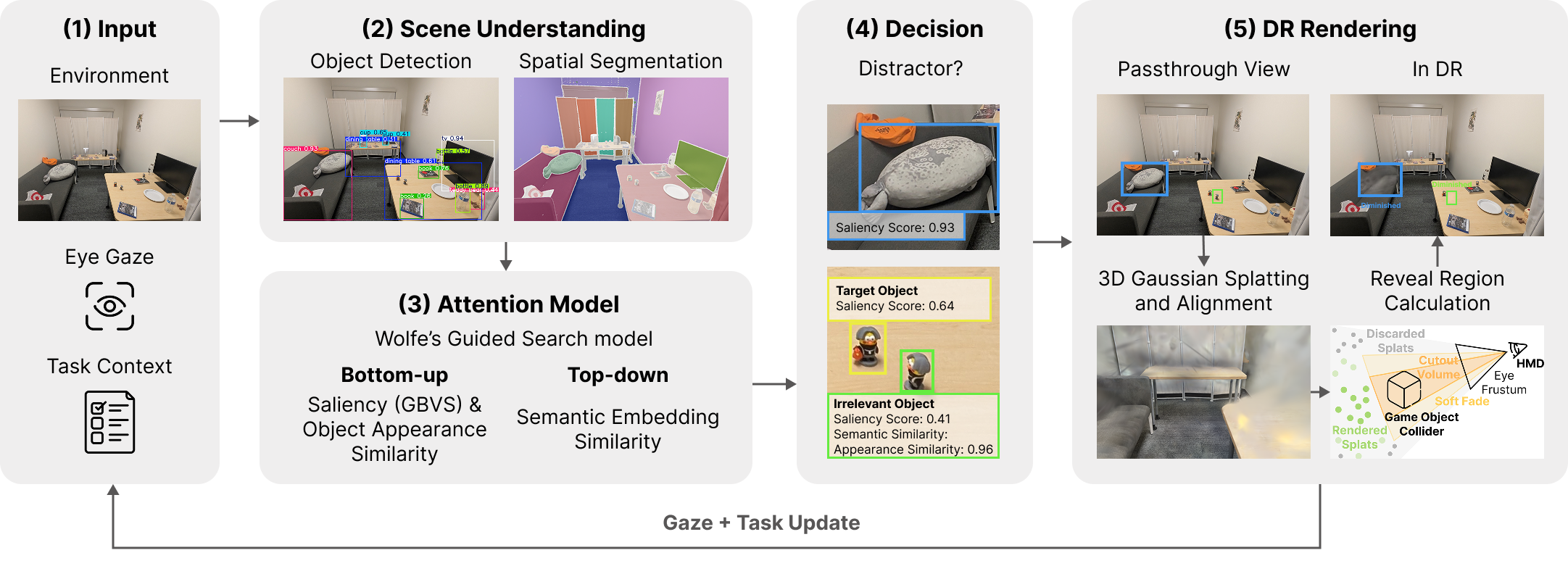}
    \caption{FocusAdapt system pipeline. In our current prototype, computationally intensive components are precomputed and replayed during the study to simulate real-time system behavior.}
    \label{fig:teaser}
  \end{center}
\end{teaserfigure}


\maketitle

\section{Introduction}


In visually cluttered procedural tasks, users must continuously focus on task-relevant objects while suppressing distractors, which may increase cognitive load and affect task efficiency. 
Emerging techniques~\cite{mann2023fundamentals, mann1999mediated} in Diminished Reality (DR) enable the real-time visual removal of irrelevant objects, thereby reducing visual clutter and supporting attentional focus in such tasks.
However, indiscriminately removing all non-target objects may disrupt task-relevant context, discard supportive environmental cues, and reduce situational awareness.
In this sense, we propose to selectively diminish distractors that significantly impact task performance, while keeping essential context present.


We first conducted a user study in which participants performed two types of procedural tasks in conditions of different visual clutter levels.
By comparing task performance and subjective experience across conditions, we found that selectively diminishing distractors reduced cognitive load as effectively as removing all task-irrelevant objects, relative to the non-diminished condition, in both tasks.
In addition, interviews with participants revealed multiple considerations for retaining task-irrelevant objects when they were not highly distracting, including their roles as contextual cues, spatial anchors, and representations of personal relevance.


Building on these results, we developed FocusAdapt, a context-aware system that selectively diminish significant distractors based attention prediction that integrates saliency analysis with task-aware semantic reasoning, inspired by Wolfe's Guided Search model~\cite{wolfe1994guided, wolfe2017five, wolfe2021guided}.
Unlike prior DR systems that rely primarily on manual selection, our system dynamically predicts distraction likelihood and selectively diminishes objects in real time.





\section{User Study}
\begin{figure}[!t]
    \centering
    \includegraphics[width=1\linewidth]{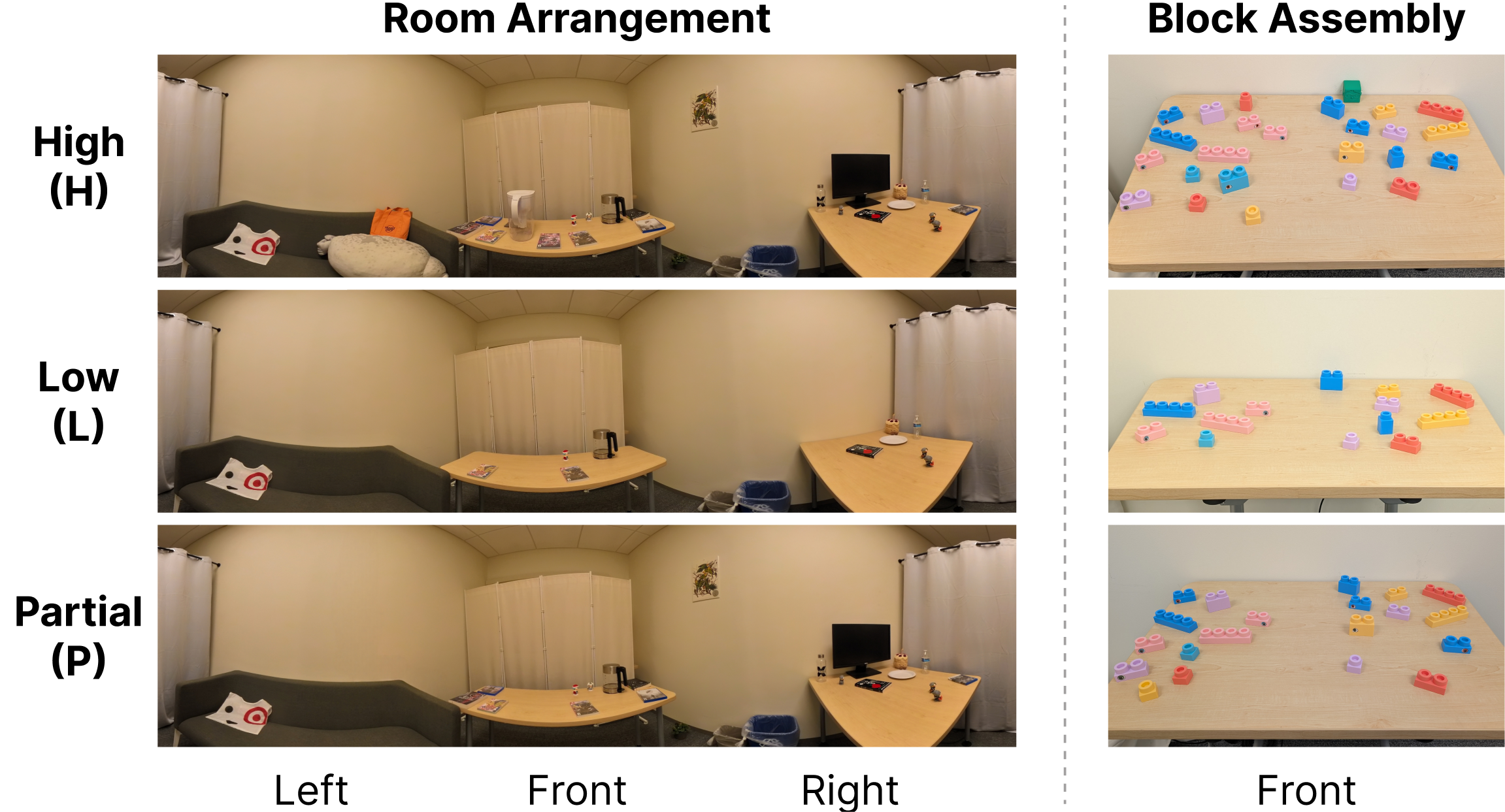}
    \caption{Formative study setting.}
    \label{fig:formative_study_setting}
    \Description{Formative Study Setting}
\end{figure}

We recruited 24 participants (Mean = 24.67 years, SD = 3.77 years; 13 male, 11 female). They were asked to complete two physically engaging procedural tasks, including \textbf{Room Organization} and \textbf{Block Assembly} in conditions of different levels of visual clutter:
\textbf{High clutter (H):} 16 task-irrelevant objects for room arrangement, and 13 task-irrelevant blocks for assembly; \textbf{Low clutter (L):} Only task-relevant objects were present; \textbf{Partial (P):} 5 task-irrelevant objects were removed from the high clutter (H) condition, selected based on participant's gazing time (Figure ~\ref{fig:formative_study_setting}). 
The order of conditions was counterbalanced using a within-subject design. 
The study was conducted in two sessions.
In the first session, participants completed four baseline conditions: Room arrangement (H), Room arrangement (L), Block assembly (H), and Block assembly (L). 
After the first session, participants took a break of at least one hour to refresh their working memory. 
The top 30\% distractors for Room Arrangement and top 40\% for Block Assembly were calculated and removed for each participant. 
In the second session, participants returned to complete two tasks in the Partial clutter condition. 
Post-task questionnaires were completed after each condition. 
Upon completing all conditions, participants took part in a semi-structured interview.

We implemented the system using Unity 6. Objects' colliders were created virtual replicas of physical objects using Polycam~\footnote{https://poly.cam/} and Blender to detect eye-gaze dwellings on the objects by analyzing fixations within the Region of Interest (ROI). We performed a standard multi-point calibration procedure \cite{holmqvist2011eye}. 
We used Meta Quest Pro for record eye tracking with a 72 Hz sampling rate \cite{hou2024unveiling}.


\subsection{Results}

The overall results show that both low clutter (L) and partial (P) conditions can reduce participants' perceived cognitive load, task completion time, and increase fixation ratio on task-relevant objects.



We analyzed questionnaire data using Friedman tests and parametric data using repeated-measures ANOVA with post hoc comparisons. Both low (L) and partial (P) clutter significantly reduced perceived mental demand ($p < 0.01$) and task completion time (room arrangement: $p < 0.01$; block assembly: $p < 0.001$) compared with high clutter (H). Fixations were identified using the I-VT algorithm \cite{salvucci2000identifying}, with fixation ratio defined as the proportion of fixation time on task-relevant objects. Compared with high clutter (H), low clutter (L) increased fixation ratios in room arrangement subtasks 6 and 8 ($p < 0.01$) and block assembly subtasks 3 ($p < 0.01$) and 4 ($p < 0.05$); partial clutter (P) also increased the ratio in room arrangement subtask 6 ($p < 0.01$).

Participants highlighted several key requirements for diminished reality systems, emphasizing that the system should be selective, context-aware, adaptive, and user-controllable rather than fully automatic or overly aggressive in removal. 
Some task-irrelevant objects should be preserved if they serve functional, contextual, cognitive, or personal roles.  


\section{FocusAdapt}

Building on insights from our user study, we developed FocusAdapt, a selective diminished reality system to assist users in completing procedural tasks.

To approximate the behavior of the envisioned real-time pipeline, computationally intensive scene-understanding components, including object detection, segmentation, and scene reconstruction, were precomputed. 
First, we perform object-level scene analysis with Ultralytics YOLO detector (yolo11s variant) for object detection \cite{jocher2023ultralytics} and SAM 2 for spatial segmentation \cite{ravi2024sam2}. To quantify perceptual similarity between objects, given an input image of an object, we extract its feature embedding using a pretrained ResNet-50 backbone~\cite{he2016deep}. 
For each object, we compute bottom-up perceptual features using a visual saliency model \cite{harel2006graph} and top-down semantic features derived from the current task step, object category, and task relevance \cite{reimers-2019-sentence-bert}.
Second, we feed these features into an attention model, which combines saliency and semantics into a predicted attention score for each object. 
This model and its thresholds are learned from our user study. At runtime, objects whose predicted attention is higher than the threshold while their task relevance is lower than the threshold are classified as distractors to be diminished, while task-relevant and other irrelevant objects are preserved.
Finally, the system passes the set of distractors to a real-time DR renderer, which diminishes them in the camera view using the 3D Gaussian Splatting technique \cite{kerbl20233d}. At runtime, FocusAdapt continuously updates attention estimates based on gaze and task context and applies the corresponding precomputed DR visualization (Figure ~\ref{fig:teaser}).

\begin{figure}[hbt!]
    \centering
    \includegraphics[width=1\linewidth]{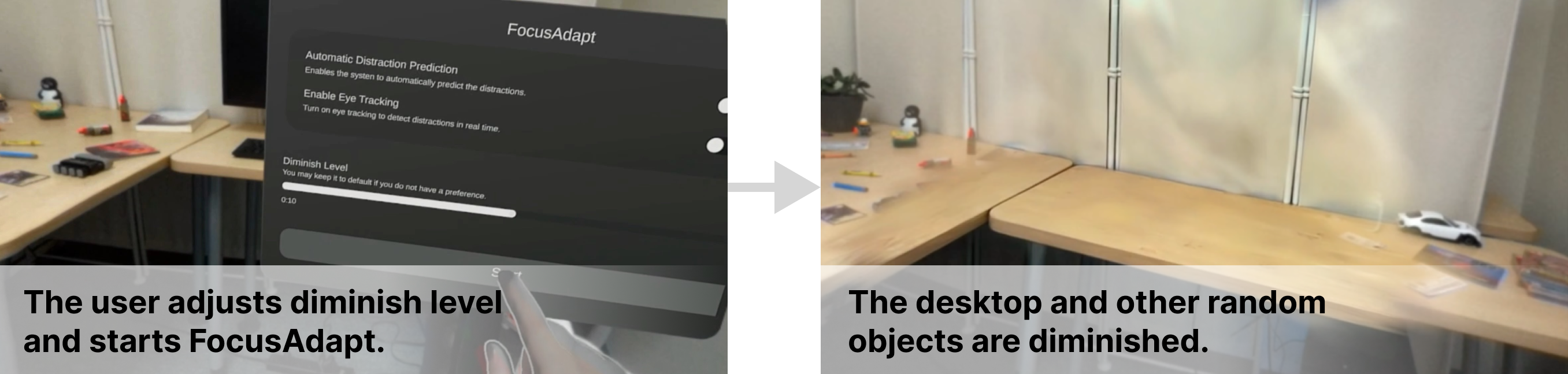}
    \caption{System demo.}
    \label{fig:system_demo}
    \Description{System Demo}
\end{figure}

Our rendering pipeline builds upon the Unity-based 3D Gaussian Splatting implementation \cite{kerbl20233d,pranckevicius_unitygaussiansplatting} (Figure ~\ref{fig:system_demo}). To reconstruct the scene, we captured the empty room without objects using approximately 10 minutes of video from an iPhone 17. We trained the Gaussian splats using Jawset Postshot (V1.0.1) on 3,000 unique frames, capping the splat count at 3 million \cite{kari2025reality}.

\vspace{-2em}
\section{Conclusion}
We presented FocusAdapt, a context-aware diminished reality system that selectively suppresses distracting objects by integrating visual saliency and similarity, semantic relevance, and user behavior. Our formative study shows that selective diminished reality reduces cognitive load, improves task efficiency, and preserves useful contextual information. These findings suggest that future DR systems should move beyond one-size-fits-all object removal toward adaptive, personalized attention support.



\bibliographystyle{ACM-Reference-Format}
\bibliography{references}

@incollection{mann2023fundamentals,
  title={Fundamentals of all the realities: Virtual, augmented, mediated, multimediated, and beyond},
  author={Mann, Steve and Do, Phillip V and Furness, Tom and Yuan, Yu and Iorio, Jay and Wang, Zixin},
  booktitle={Springer handbook of augmented reality},
  pages={3--34},
  year={2023},
  publisher={Springer}
}

@article{mann1999mediated,
author = {Mann, Steve},
title = {Mediated Reality},
year = {1999},
issue_date = {March 1999},
publisher = {Belltown Media},
address = {Houston, TX},
volume = {1999},
number = {59es},
issn = {1075-3583},
journal = {Linux J.},
month = mar,
pages = {5–es}
}

@article{wolfe2017five,
  title={Five factors that guide attention in visual search},
  author={Wolfe, Jeremy M and Horowitz, Todd S},
  journal={Nature human behaviour},
  volume={1},
  number={3},
  pages={0058},
  year={2017},
  publisher={Nature Publishing Group UK London}
}

@article{wolfe2021guided,
  title={Guided Search 6.0: An updated model of visual search},
  author={Wolfe, Jeremy M},
  journal={Psychonomic bulletin \& review},
  volume={28},
  number={4},
  pages={1060--1092},
  year={2021},
  publisher={Springer}
}

@article{wolfe1994guided,
  title={Guided search 2.0 a revised model of visual search},
  author={Wolfe, Jeremy M},
  journal={Psychonomic bulletin \& review},
  volume={1},
  number={2},
  pages={202--238},
  year={1994},
  publisher={Springer}
}

@inproceedings{kari2025reality,
  title={Reality Promises: Virtual-Physical Decoupling Illusions in Mixed Reality via Invisible Mobile Robots},
  author={Kari, Mohamed and Abtahi, Parastoo},
  booktitle={Proceedings of the 38th Annual ACM Symposium on User Interface Software and Technology},
  pages={1--17},
  year={2025}
}

@article{harel2006graph,
  title={Graph-based visual saliency},
  author={Harel, Jonathan and Koch, Christof and Perona, Pietro},
  journal={Advances in neural information processing systems},
  volume={19},
  year={2006}
}

@inproceedings{salvucci2000identifying,
  title={Identifying fixations and saccades in eye-tracking protocols},
  author={Salvucci, Dario D and Goldberg, Joseph H},
  booktitle={Proceedings of the 2000 symposium on Eye tracking research \& applications},
  pages={71--78},
  year={2000}
}

@inproceedings{hou2024unveiling,
  title={Unveiling variations: A comparative study of vr headsets regarding eye tracking volume, gaze accuracy, and precision},
  author={Hou, Baosheng James and Abdrabou, Yasmeen and Weidner, Florian and Gellersen, Hans},
  booktitle={2024 IEEE Conference on Virtual Reality and 3D User Interfaces Abstracts and Workshops (VRW)},
  pages={650--655},
  year={2024},
  organization={IEEE}
}

@inproceedings{he2016deep,
  title={Deep residual learning for image recognition},
  author={He, Kaiming and Zhang, Xiangyu and Ren, Shaoqing and Sun, Jian},
  booktitle={Proceedings of the IEEE conference on computer vision and pattern recognition},
  pages={770--778},
  year={2016}
}

@book{holmqvist2011eye,
  title={Eye tracking: A comprehensive guide to methods and measures},
  author={Holmqvist, Kenneth and Nystr{\"o}m, Marcus and Andersson, Richard and Dewhurst, Richard and Jarodzka, Halszka and Van de Weijer, Joost},
  year={2011},
  publisher={oup Oxford}
}

@article{ravi2024sam2,
  title={SAM 2: Segment Anything in Images and Videos},
  author={Ravi, Nikhila and Gabeur, Valentin and Hu, Yuan-Ting and Hu, Ronghang and Ryali, Chaitanya and Ma, Tengyu and Khedr, Haitham and R{\"a}dle, Roman and Rolland, Chloe and Gustafson, Laura and Mintun, Eric and Pan, Junting and Alwala, Kalyan Vasudev and Carion, Nicolas and Wu, Chao-Yuan and Girshick, Ross and Doll{\'a}r, Piotr and Feichtenhofer, Christoph},
  journal={arXiv preprint arXiv:2408.00714},
  url={https://arxiv.org/abs/2408.00714},
  year={2024}
}

@article{jocher2023ultralytics,
  title={Ultralytics YOLOv8},
  author={Jocher, Glenn and others},
  journal={GitHub repository},
  year={2023},
  howpublished={\url{https://github.com/ultralytics/ultralytics}}
}

@misc{pranckevicius_unitygaussiansplatting,
  author       = {Aras Pranckevi\v{c}ius},
  title        = {UnityGaussianSplatting},
  howpublished = {\url{https://github.com/aras-p/UnityGaussianSplatting}},
  note         = {GitHub repository, accessed 2026-03-29},
  year         = {2026}
}

@article{kerbl20233d,
  title={3d gaussian splatting for real-time radiance field rendering.},
  author={Kerbl, Bernhard and Kopanas, Georgios and Leimk{\"u}hler, Thomas and Drettakis, George and others},
  journal={ACM Trans. Graph.},
  volume={42},
  number={4},
  pages={139--1},
  year={2023}
}

@inproceedings{reimers-2019-sentence-bert,
    title = "Sentence-BERT: Sentence Embeddings using Siamese BERT-Networks",
    author = "Reimers, Nils and Gurevych, Iryna",
    booktitle = "Proceedings of the 2019 Conference on Empirical Methods in Natural Language Processing and the 9th International Joint Conference on Natural Language Processing (EMNLP-IJCNLP)",
    year = "2019",
    publisher = "Association for Computational Linguistics",
}


\end{document}